\documentclass{sigchi-ext}
\usepackage[T1]{fontenc}
\usepackage{textcomp}
\usepackage[scaled=.92]{helvet} 
\usepackage{graphicx} 
\usepackage{balance}  
\usepackage{booktabs} 
\usepackage{ccicons}  
\usepackage{ragged2e} 

\copyrightinfo{
\newline \newline
}

\def\plaintitle{SIGCHI Extended Abstracts Sample File: Note Initial
  Caps} 
\def\emptyauthor{}
\def\plainkeywords{Long-Term Self-Tracking; Health; Physical Activity; Longitudinal study; Apple HealthKit}

\title{Activity Self-Tracking with Smart Phones: How to Approach Odd Measurements?}

\numberofauthors{2}
\author{%
  \alignauthor{%
    \textbf{Gabriela Villalobos-Z\'u\~niga}\\
    \affaddr{University of Lausanne} \\
    \email{gabriela.villalobos@unil.ch} }\alignauthor{%
    \textbf{Mauro Cherubini}\\
    \affaddr{University of Lausanne}\\
    \email{mauro.cherubini@unil.ch} } }

\teaser{
  \includegraphics[width=0.8\columnwidth]{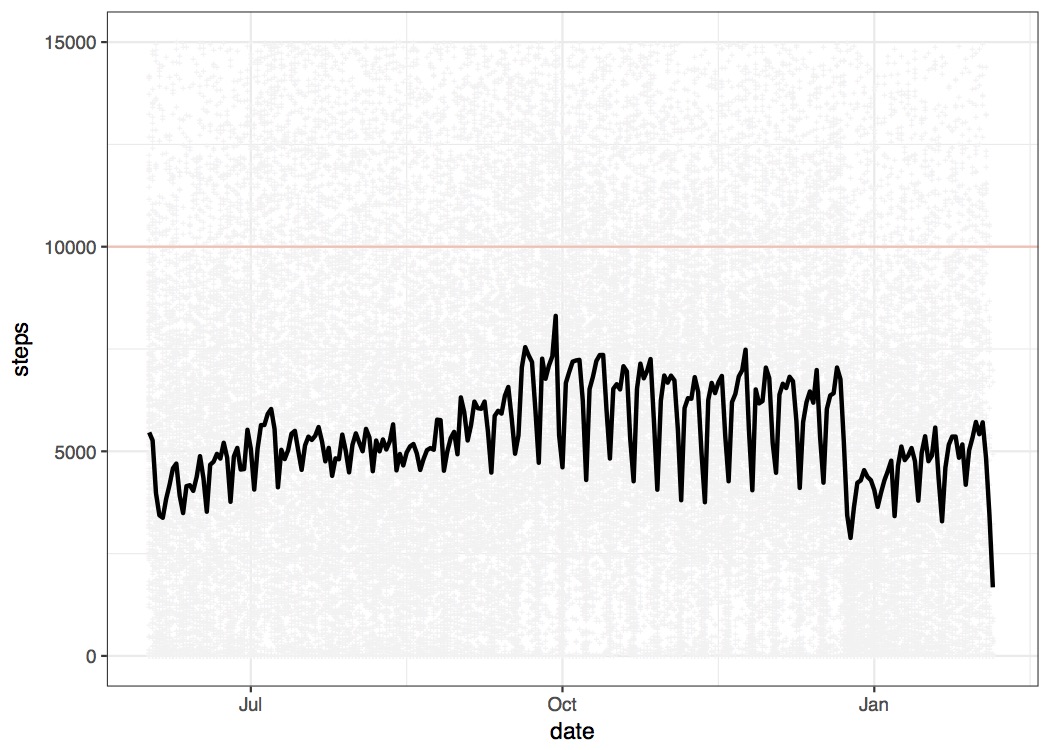}
  \caption{Seasonality Effect - General Curve of Means. The "pulse" regularity is an indication of seasonality effects.}
  \label{fig:seasonality}
}    

\definecolor{linkColor}{RGB}{6,125,233}
\hypersetup{%
  pdftitle={\plaintitle},
  pdfauthor={\emptyauthor},
  pdfkeywords={\plainkeywords},
  bookmarksnumbered,
  pdfstartview={FitH},
  colorlinks,
  citecolor=black,
  filecolor=black,
  linkcolor=black,
  urlcolor=linkColor,
  breaklinks=true,
}

\begin{document}


\maketitle
\RaggedRight{}

\begin{abstract}
  Tracking physical activity reliably is becoming central to many research efforts. In the last years specialized hardware has been proposed to measure movement. However, asking study participants to carry additional devices has drawbacks. We focus on using mobile devices as motion sensors. In the paper we detail several issues that we found while using this technique in a longitudinal study involving hundreds of participants for several months. We hope to sparkle a lively discussion at the workshop and attract interest in this method from other researchers.
\end{abstract}

\keywords{\plainkeywords}

\category{H.5.m}{Information interfaces and presentation (e.g.,
  HCI)}{Miscellaneous}

\section{Introduction}
Physical inactivity is the fourth leading risk factor for global mortality \footnote{http://www.who.int/topics/physical\_activity/en/,last retrieve February 2018.}. Diseases such as coronary heart disease, type 2 diabetes, breast and colon cancers are caused by insufficient physical activity. Furthermore, the lack of moderate intensity physical activity, is responsible for 9\% of premature mortality and 3.2 million deaths globally \cite{lee2012effect}. Due to the overwhelming scientific evidence on the benefits of physical activity \cite{warburton2006health}, it becomes clear the necessity to implement mechanisms to increase the physical activity levels worldwide. Guidelines recommend adults to practice a moderate-intensity physical activity for at least 30 minutes a day during 5 days a week \cite{garber2011american}.
%
%

The points raised before justify the growing body of research aiming to analyze, support and/or enhance human activities through the means of technology \cite{friederichs2014move},\cite{boudreau2016effectiveness},\cite{hurling2007using},\cite{nes2017personalized}. These research efforts focus on diverse populations: diabetics \cite{bonn2018app}, cancer survivors \cite{robertson2017mobile},\cite{haggerty2017randomized}, children \cite{pakarinen2017health}, etc. being their common goal to attain the adoption and/or maintenance of physical activity. Furthermore, industry is as well engaged in the implementation of technology supported means to stimulate physical activity. Examples of these include, fitness coaching apps\footnote{E.g., https://www.skimble.com/, last retrieved February 2018.}, and run tracking applications\footnote{Eg., https://www.runtastic.com/, last retrieved February 2018.}.  



From a hardware standpoint, there has been a growing number of specialized devices that have been developed to sense human activity. These include bracelets, clips\footnote{Eg., https://www.fitbit.com/zip, last retrieved February 2018} or wearable sensors\footnote{Eg., https://www.dexcom.com/continuous-glucose-monitoring, last retrieved February 2018.}. While as consumer products these devices are great, as research devices they suffer serious limitations.  For instance, users might forget to wear them consistently, or their battery could run out of power if not systematically recharged. These points --among others-- persuaded us to reconsider employing just smart phones without any additional hardware as a tool for tracking steps.
%
%

\begin{margintable}[-23pc]
  \begin{minipage}{\marginparwidth}
    \centering
    \begin{tabular}{r l}
    	{\small \textbf{Statistic}}
      & {\small \textbf{Value}}
      \\
      \toprule
      Min. & 0 \\
      1st Qu. & 2765\\
      Median & 5584 \\
      Mean & 6490\\
      3rd Qu. & 8916\\
      Max. & 98065 \\
      NA's & 1760 \\
 
    \end{tabular}
    \caption{Descriptive statistics of the step count distribution.}~\label{tab:table1}
  \end{minipage}
\end{margintable}

\section{Using Smart Phones To Track Steps}
An elegant solution to these limitations is that of using smart phones as sensors of physical activity. These can accurately predict walking activity \cite{nolan2014validity} and have been previously used to measure physical exercise \cite{liu2016exercise},\cite{harries2016effectiveness}.  Moreover, they have the following advantages: they act as \textit{silent observers}, letting participants carry on with their tasks without an explicit reminder of being tracked, thus making the data capturing less intrusive. Furthermore, people are willing to carry their phones with them during the majority of their daily activities and are also aware of the level of battery these have because they want to remain active in their social networks (e.g., Facebook, Snapchat, Twitter, Instagram) and reachable to instant messaging and calls  (e.g., Whatsapp, Skype, SMS). The fact that most of day-time period users carry their phones, diminishes the negative factors that come naturally with the usage of trackers where users forget to charge it or wear it. Finally, another methodological advantage of using the smart phone as a sensor is that it does not require participants to use additional hardware that they usually would not, thus increasing the ecological validity of the study.

Of the different hardware solutions, we focus on the iPhone because Apple has standardized both the hardware and the API with which we can collect activity data, furthermore its use diminishes the development costs required to build the experimental infrastructure for a study. 

Apple's HealthKit\textsuperscript{TM} is a platform with a repository of physical activity data collected from the iPhone's accelerometer and other health data obtained from various sensors such as scales and blood testing devices. Data collected is stored in an encrypted database called \textit{Healthkit Store} from which steps are retrieved\footnote{HealthKit also collects data from a variety of other sensors. However, in the context of this paper we focus only on steps measurements.}. Researchers have made use of this platform in the past to implement steps tracking applications \cite{frank2017personalized},\cite{jeong2017smartwatch},\cite{xu2017analysis}. 

As part of ongoing research we used Apple's HealthKit\textsuperscript{TM} platform as a sensor to collect physical activity data. The goal of this paper is to discuss advantages and disadvantages of this method, and to report specific issues we encountered and possible solutions. We plan to contribute to the workshop by sparkling interest in this method and discussion on how to address some of the issues that we encountered.

\section{Adversities of Tracking Steps With a Smart Phone}
We are currently using Apple's Health Kit\textsuperscript{TM} to record the physical activity of users as part of a research project. The study involves an longitudinal observation period of 6 months. 230 students are currently participating in this research(62\% female and 38\% male), average age of 21(SD = 2.3). Data capture started in June 2017 and it is still ongoing. While analyzing the data, we observed a number of issues that we would like to discuss here.

\begin{marginfigure}[-29pc]
  \begin{minipage}{\marginparwidth}
    \centering
    \includegraphics[width=0.9\marginparwidth]{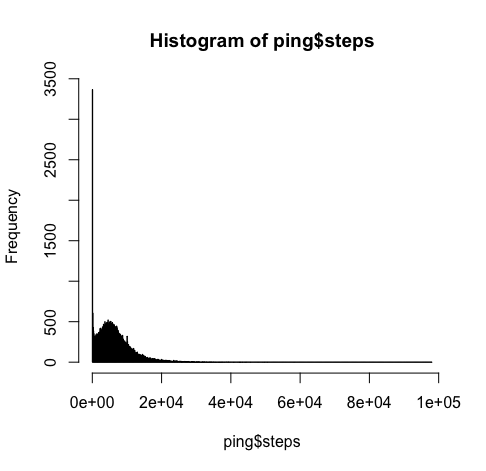}
    \caption{Step Count Distribution. The distribution is left skewed due to the presence of small steps measurements returned by the smart phone.} \label{fig:histograms1}
  \end{minipage}
\end{marginfigure}

\begin{marginfigure}[1pc]
  \begin{minipage}{\marginparwidth}
    \centering
    \includegraphics[width=0.9\marginparwidth]{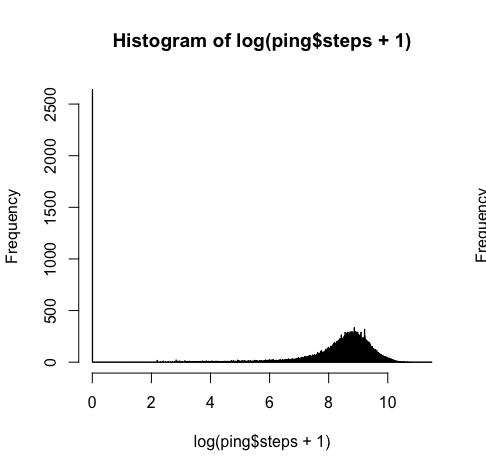}
    \caption{Log-transform of the step count distribution. While the small values are still visible, the distribution assumes a more symmetrical shape.} \label{fig:histograms2}
  \end{minipage}
\end{marginfigure}

\textit{Unreliability of the sensor to very low physical activity}: We observed a conspicuous number of days for which the activity sensors provided unreliable data. These are days which correspond to extremely low physical activity of the participants. During those days, the sensor was unable to provide an accurate measure of the activity of the participants providing counts equal to zero or little steps or returning a NA (i.e., not available) altogether. Obviously, this does not correspond to reality as we interviewed participants and indeed they moved during these days. See Table \ref{tab:table1} for descriptive statistics of the steps distribution. We are still unsure of why this happens. This could be due to the inability of the internal accelerometer and its corresponding signal processing algorithm to distinguish human steps when the pace is shorter or slower than usual (e.g., moving indoor vs. outdoor).
\textbf{Solution}: Our short-term solution to this problem is that of discarding low activity values. To do that we plotted a histogram of the collected steps and noticed skewness towards zero of the distribution curve. We took the logarithmic-transform of the curve and noticed that to make it symmetrical we should remove measurements below 400 steps. See Figures \ref{fig:histograms1} and \ref{fig:histograms2} for context.

\begin{marginfigure}[-0.5pc]
  \begin{minipage}{\marginparwidth}
    \centering
    \includegraphics[width=0.9\marginparwidth]{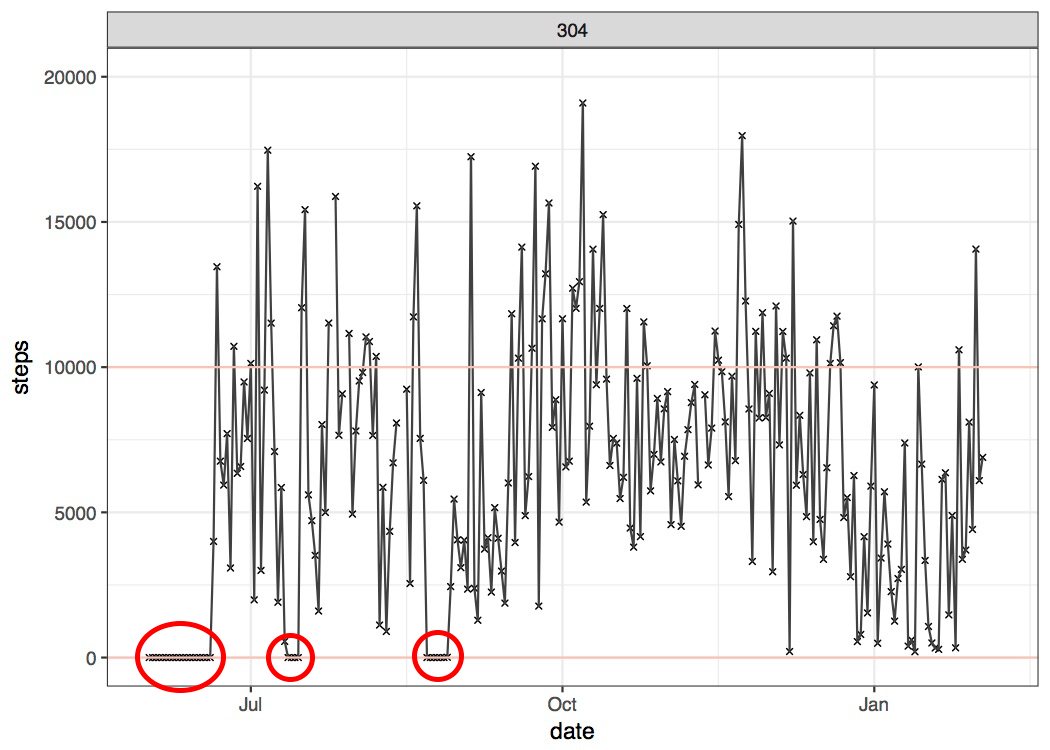}
    \caption{Example of micro-holes in the longitudinal measurements (circled in red). The number on the top of the graph represent the participant id.} \label{fig:micro-holes}
  \end{minipage}
\end{marginfigure}

\textit{Health App configuration changes}: In order for Health Kit\textsuperscript{TM} to record steps, the user needs to grant permission to access sensor data. This is is verified by turning on the \textit{Fitness Tracking} switch on the iPhone menu \textit{Settings->Privacy->Motion \& Fitness}. In our study we noticed periods in which despite of receiving normal data since the beginning of the study we suddenly started obtaining zeros, see Figure \ref{fig:micro-holes}. When contacting participants and verifying their iPhone configuration we noticed that the Fitness Tracking switch was turned off, resulting in the inability of the sensor to register steps. Another similar issue occurred when the participant did not grant our research application the rights to access Health data, from which we retrieve the number of steps. \textbf{Solution}: Our practical solution was to contact directly these participants and follow a step by step procedure to verify that all the configuration parameters were set up accordingly.  We also implemented a feature in the research application that checked the settings whenever the application was opened. If these were turned off, the user was redirected to the \textit{Motion \& Fitness} settings to turn the switch back on.

\textit{Micro-holes in the dataset}: We observed an evident number of days in which the count of steps was reported zero. This occurs randomly across the dataset. See Figure \ref{fig:micro-holes}. We believe this appears due to a technical error in the sensor that prevents it from recording steps.
\textbf{Solution}: Our solution approach is the same as the one employed for unreliability of the sensor to very low physical activity.

\begin{marginfigure}[2pc]
  \begin{minipage}{\marginparwidth}
    \centering
    \includegraphics[width=0.9\marginparwidth]{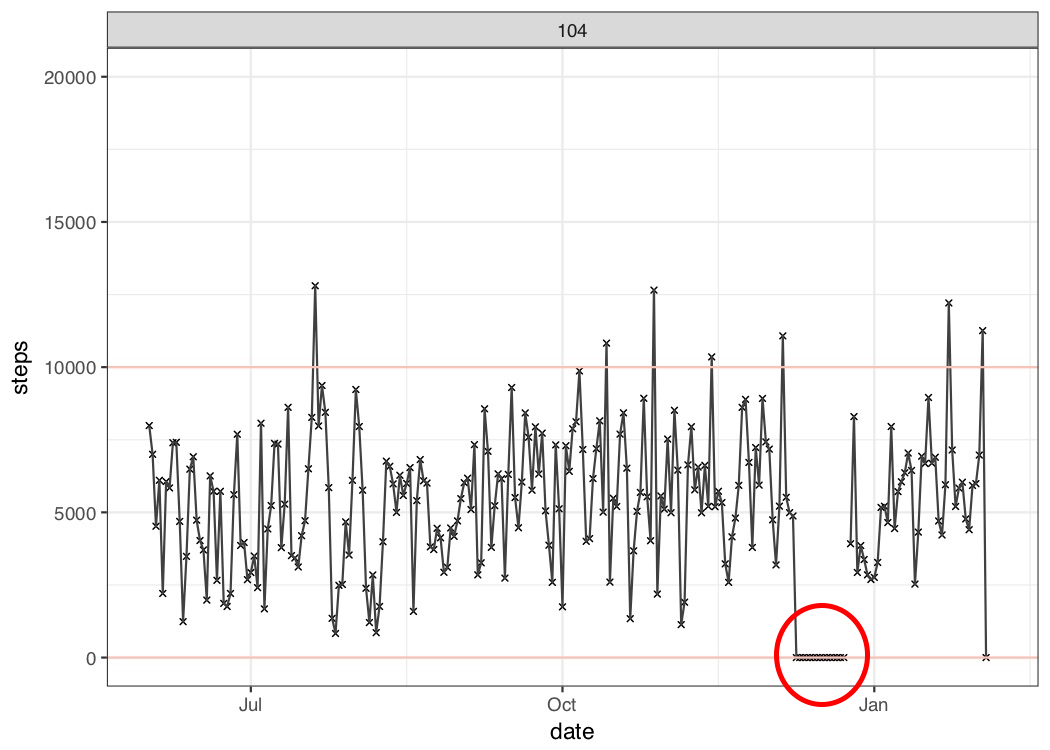}
    \caption{Example of macro-holes.} \label{fig:macro-holes1}
  \end{minipage}
\end{marginfigure}

\begin{marginfigure}[3pc]
  \begin{minipage}{\marginparwidth}
    \centering
    \includegraphics[width=0.9\marginparwidth]{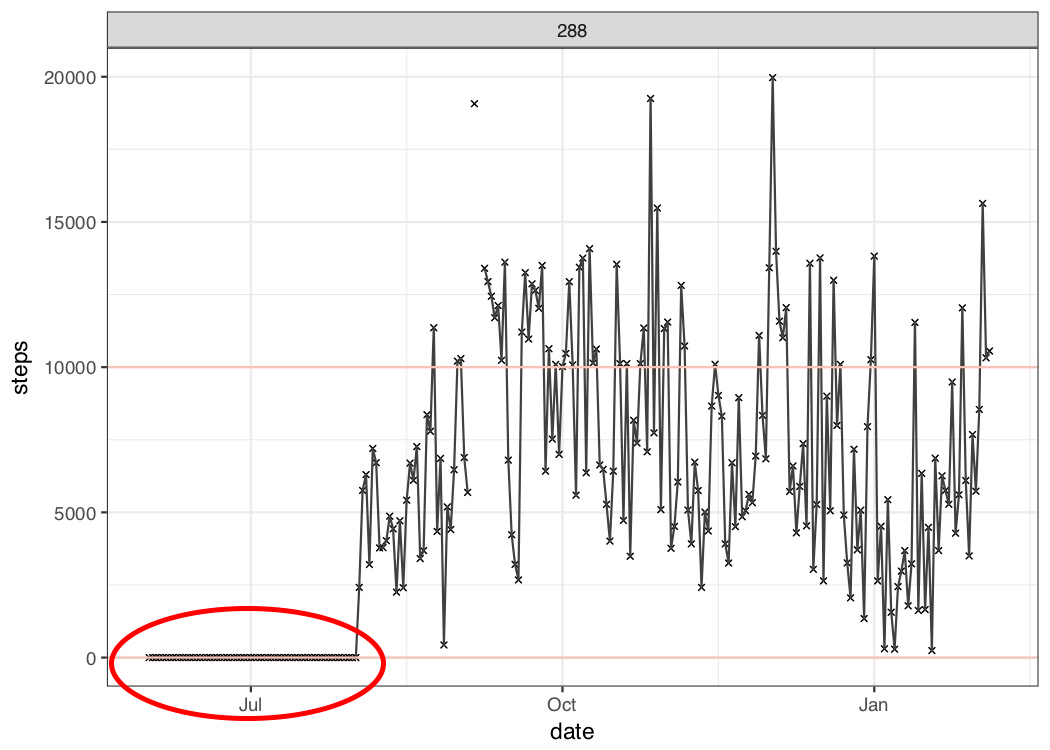}
    \caption{Additional example of macro-holes.} \label{fig:macro-holes2}
  \end{minipage}
\end{marginfigure}

\textit{Macro-holes in the dataset}: There is no activity recorded for several consecutive days. We believe that these 'macro-holes' are generated in two distinct situations: in some cases, we observe continuous zeros for some participants at the onset of the study. The justification for this is that the research application was not yet installed into the participant's device, therefore no permission to access sensor data was granted. In some other cases, if the macro-hole is found in the middle of the study, this might be caused by the participant who turned off the permission for our app to access the sensor data. See Figures \ref{fig:macro-holes1} and \ref{fig:macro-holes2}.
\textbf{Solution}: The approach here is to use a statistical analysis method that is able to deal with missing data. We chose to use Linear Mixed-Effects Regression (or LMER). Each subject in these LMER models may vary in terms of the number of measurement occasions. Subjects who are missing data at a given time point are not excluded from the analysis. In considering missing data and whether they are ignorable or not, a related issue is the distinction between \textit{attrition} (i.e., subjects dropping out of the study and not returning) and \textit{sporadic} or \textit{intermittent data} (i.e., subjects with missing data between observed time-points). Attrition should be consider carefully and might not be ignored. Participants where consistent attrition is observed should be removed from the analysis.

\begin{marginfigure}[-6pc]
  \begin{minipage}{\marginparwidth}
    \centering
    \includegraphics[width=0.9\marginparwidth]{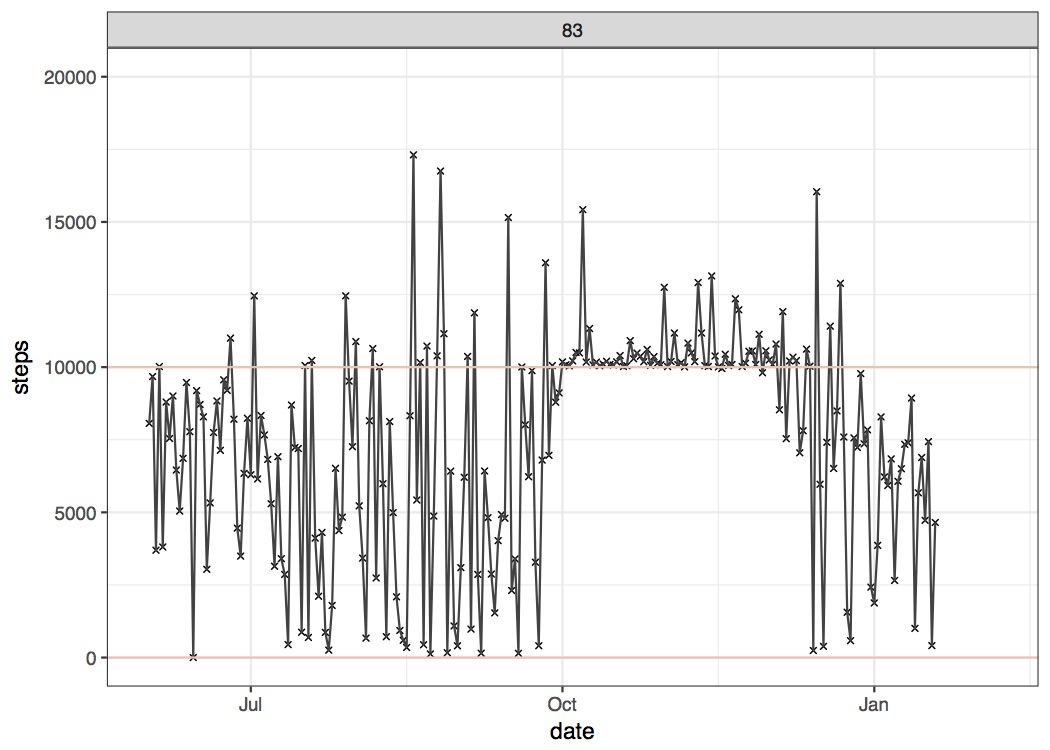}
    \caption{Plateau effect, as observed in one of the participants.} \label{fig:plateau}
  \end{minipage}
\end{marginfigure}

\textit{Plateau in the dataset}: We noticed data for which in a particular moment in time augmented until reaching above 10'000 steps a day and remained constant for some months. We believe participants in this situation were trying to reach the daily step goal we set to 10'000 steps.  See Figure \ref{fig:plateau}.
\textbf{Solution}: It is still not clear in our analysis whether these participants cheated by tampering the sensor or whether they actually had physical activity consistently above the set threshold of 10K steps for several weeks in a row. To counter people who wanted to cheat in the experiment, we explicitly excluded manually entered steps from the step count. However, participants might have found new creative ways to cheat that we have not identified yet.

\textit{Seasonality effects}: Physical activity of the participants was influenced by the summer holiday season (July and August). Period in which participants tend not to carry their phone or they perform more or less exercise than their normal average. Similarly, occurs during weekdays and weekends in which people tend to be less active in the later one. These effect creates uncertainty in the effects of the treatment of the study by not allowing to clearly determine what increased or decreased the physical activity.  See Figure \ref{fig:seasonality}.
\textbf{Solution}: Our solution for this case consists in removing the seasonality effects by putting in practice methods borrowed from time-series analysis such as effect decomposition.  

\section{Discussion}
Determining the correct type of devices to track data for research studies is critical for their validity. This decision might depend on the design of the experiment and the specific research questions the researchers are tackling. In some cases it might be adequate to ask participants to carry additional hardware, in others it might not. We argue that even with the limitations presented in this position paper, using smart phones as sensor of physical activity is a powerful mechanism to study human activity and a valid research method.
There are a number of open questions that we would like to ask the audience and get feedback on during the workshop.

\textit{Which method should we use to process micro-holes in the dataset?} There are two possible solutions to this: either we use the last-value carried forward (or LVCF) or interpolation between two known measurements. While the former is more respectful of the auto-regressive nature of longitudinal data \cite{genccay2001introduction}, the latter is not. We would like to hear expert opinions on this issue.

\textit{What are the possible explanations for NAs being returned from Apple HealthKit\textsuperscript{TM}?} We know little of the inner working of the sensor and the built-in algorithm that process the accelerometer signals in order to understand the fringe cases that we observed. We would like to ask the audience whether any technical documentation is available on this specific hardware or any engineer who might have reverse-engineered the inner working of this black box.

\textit{Removing seasonality effects or keeping them in the data?} We are still unsure whether the best method to analyze longitudinal data is that of removing the seasonality effects. Contrary to time series, where measurements are said to be independent, data coming from longitudinal studies have an auto-regressive nature that we should not ignore.  Seasonality is certainly part of how people behave: activity on week days might be consistently lower for many users. Should we remove this effects?

\textit{How to best exploit iPhone background execution mode for experimental purposes?} Apple allows background mode for specific purposes such as: playing audio, receiving location updates, performing finite-length task, and background fetch (e.g. news retrieval). Unfortunately none of those fit out need, and this forced us to ask our participants to open the application periodically (bring it to foreground mode) so we could retrieve the steps from HealthKit\textsuperscript{TM} . Avoiding intervening so directly into the experiment, would increase the ecological validity of the study. Recently, we found a mechanism in which by using the option : \textit{UIApplicationBackgroundFetchIntervalMinimum} we are able to overcome this issue. Nonetheless, we would love to discuss whether other solutions to this problematic have been exploited successfully. 

\section{Conclusion and Next Steps}
Having clear/reliable data is central to long-term self tracking, hence the importance of working with precise and solid measure instruments. Researchers can make use of the variety of trackers that are offered by the market, however all existing solutions have drawbacks. With the purpose of obtaining valid results, it is imperative that researchers find workarounds to standing issues, making the discussion of these problems highly relevant for the research community.

In a follow-up study we are planning to improve the quality of the data we capture working on the different aspects discussed in this paper.  

\balance{} 

\bibliographystyle{SIGCHI-Reference-Format}
\bibliography{sample}


\begin{thebibliography}{00}


\ifx \showCODEN    \undefined \def \showCODEN     #1{\unskip}     \fi
\ifx \showDOI      \undefined \def \showDOI       #1{{\tt DOI:}\penalty0{#1}\ }
  \fi
\ifx \showISBNx    \undefined \def \showISBNx     #1{\unskip}     \fi
\ifx \showISBNxiii \undefined \def \showISBNxiii  #1{\unskip}     \fi
\ifx \showISSN     \undefined \def \showISSN      #1{\unskip}     \fi
\ifx \showLCCN     \undefined \def \showLCCN      #1{\unskip}     \fi
\ifx \shownote     \undefined \def \shownote      #1{#1}          \fi
\ifx \showarticletitle \undefined \def \showarticletitle #1{#1}   \fi
\ifx \showURL      \undefined \def \showURL       #1{#1}          \fi

\bibitem{bonn2018app}
{Stephanie~E Bonn}, {Christina Alexandrou}, {Kristin~Hj{\"o}rleifsdottir
  Steiner}, {Klara Wiklander}, {Claes-G{\"o}ran {\"O}stenson}, {Marie L{\"o}f},
  {and} {Ylva~Trolle Lagerros}. 2018.
\newblock \showarticletitle{App-technology to increase physical activity among
  patients with diabetes type 2-the DiaCert-study, a randomized controlled
  trial}.
\newblock {\em BMC public health\/} {18}, 1 (2018), 119.
\newblock


\bibitem{boudreau2016effectiveness}
{Fran{\c{c}}ois Boudreau}, {Michel Moreau}, {and} {Jos{\'e} C{\^o}t{\'e}}.
  2016.
\newblock \showarticletitle{Effectiveness of computer tailoring versus peer
  support web-based interventions in promoting physical activity among
  insufficiently active Canadian adults with type 2 diabetes: protocol for a
  randomized controlled trial}.
\newblock {\em JMIR research protocols\/} {5}, 1 (2016).
\newblock


\bibitem{frank2017personalized}
{Julian Frank}. 2017.
\newblock {\em A personalized support tool for the training of mindful walking:
  The mobile 'MindfulWalk' application}.
\newblock Ph.D. Dissertation. Ulm University.
\newblock


\bibitem{friederichs2014move}
{Stijn~AH Friederichs}, {Anke Oenema}, {Catherine Bolman}, {Janneke Guyaux},
  {Hilde~M van Keulen}, {and} {Lilian Lechner}. 2014.
\newblock \showarticletitle{I Move: systematic development of a web-based
  computer tailored physical activity intervention, based on motivational
  interviewing and self-determination theory}.
\newblock {\em BMC public health\/} {14}, 1 (2014), 212.
\newblock


\bibitem{garber2011american}
{Carol~Ewing Garber}, {Bryan Blissmer}, {Michael~R Deschenes}, {Barry~A
  Franklin}, {Michael~J Lamonte}, {I-Min Lee}, {David~C Nieman}, {and} {David~P
  Swain}. 2011.
\newblock \showarticletitle{American College of Sports Medicine position stand.
  Quantity and quality of exercise for developing and maintaining
  cardiorespiratory, musculoskeletal, and neuromotor fitness in apparently
  healthy adults: guidance for prescribing exercise.}
\newblock {\em Medicine and science in sports and exercise\/} {43}, 7 (2011),
  1334--1359.
\newblock


\bibitem{genccay2001introduction}
{Ramazan Gen{\c{c}}ay}, {Michel Dacorogna}, {Ulrich~A Muller}, {Olivier
  Pictet}, {and} {Richard Olsen}. 2001.
\newblock {\em An introduction to high-frequency finance}.
\newblock Elsevier.
\newblock


\bibitem{haggerty2017randomized}
{Ashley~F Haggerty}, {Andrea Hagemann}, {Matthew Barnett}, {Mark Thornquist},
  {Marian~L Neuhouser}, {Neil Horowitz}, {Graham~A Colditz}, {David~B Sarwer},
  {Emily~M Ko}, {and} {Kelly~C Allison}. 2017.
\newblock \showarticletitle{A Randomized, Controlled, Multicenter Study of
  Technology-Based Weight Loss Interventions among Endometrial Cancer
  Survivors}.
\newblock {\em Obesity\/} {25}, S2 (2017).
\newblock


\bibitem{harries2016effectiveness}
{Tim Harries}, {Parisa Eslambolchilar}, {Ruth Rettie}, {Chris Stride}, {Simon
  Walton}, {and} {Hugo~C van Woerden}. 2016.
\newblock \showarticletitle{Effectiveness of a smartphone app in increasing
  physical activity amongst male adults: a randomised controlled trial}.
\newblock {\em BMC public health\/} {16}, 1 (2016), 925.
\newblock


\bibitem{hurling2007using}
{Robert Hurling}, {Michael Catt}, {Marco De~Boni}, {Bruce~William Fairley},
  {Tina Hurst}, {Peter Murray}, {Alannah Richardson}, {and} {Jaspreet~Singh
  Sodhi}. 2007.
\newblock \showarticletitle{Using internet and mobile phone technology to
  deliver an automated physical activity program: randomized controlled trial}.
\newblock {\em Journal of medical Internet research\/} {9}, 2 (2007).
\newblock


\bibitem{jeong2017smartwatch}
{Hayeon Jeong}, {Heepyung Kim}, {Rihun Kim}, {Uichin Lee}, {and} {Yong Jeong}.
  2017.
\newblock \showarticletitle{Smartwatch Wearing Behavior Analysis: A
  Longitudinal Study}.
\newblock {\em Proceedings of the ACM on Interactive, Mobile, Wearable and
  Ubiquitous Technologies\/} {1}, 3 (2017), 60.
\newblock


\bibitem{lee2012effect}
{I-Min Lee}, {Eric~J Shiroma}, {Felipe Lobelo}, {Pekka Puska}, {Steven~N
  Blair}, {Peter~T Katzmarzyk}, {Lancet Physical Activity Series~Working
  Group}, {and} {others}. 2012.
\newblock \showarticletitle{Effect of physical inactivity on major
  non-communicable diseases worldwide: an analysis of burden of disease and
  life expectancy}.
\newblock {\em The lancet\/} {380}, 9838 (2012), 219--229.
\newblock


\bibitem{liu2016exercise}
{Chung-Tse Liu} {and} {Chia-Tai Chan}. 2016.
\newblock \showarticletitle{Exercise Performance Measurement with Smartphone
  Embedded Sensor for Well-Being Management}.
\newblock {\em International journal of environmental research and public
  health\/} {13}, 10 (2016), 1001.
\newblock


\bibitem{nes2017personalized}
{Bjarne~M Nes}, {Christian~R Gutvik}, {Carl~J Lavie}, {Javaid Nauman}, {and}
  {Ulrik Wisl{\o}ff}. 2017.
\newblock \showarticletitle{Personalized activity intelligence (PAI) for
  prevention of cardiovascular disease and promotion of physical activity}.
\newblock {\em The American journal of medicine\/} {130}, 3 (2017), 328--336.
\newblock


\bibitem{nolan2014validity}
{Meaghan Nolan}, {J~Ross Mitchell}, {and} {Patricia~K Doyle-Baker}. 2014.
\newblock \showarticletitle{Validity of the Apple iPhone{\textregistered}/iPod
  Touch{\textregistered} as an accelerometer-based physical activity monitor: a
  proof-of-concept study}.
\newblock {\em Journal of Physical Activity and Health\/} {11}, 4 (2014),
  759--769.
\newblock


\bibitem{pakarinen2017health}
{Anni Pakarinen}, {Heidi Parisod}, {Jouni Smed}, {and} {Sanna Salanter{\"a}}.
  2017.
\newblock \showarticletitle{Health game interventions to enhance physical
  activity self-efficacy of children: a quantitative systematic review}.
\newblock {\em Journal of advanced nursing\/} {73}, 4 (2017), 794--811.
\newblock


\bibitem{robertson2017mobile}
{Michael~C Robertson}, {Edward Tsai}, {Elizabeth~J Lyons}, {Sanjana
  Srinivasan}, {Maria~C Swartz}, {Miranda~L Baum}, {and} {Karen~M
  Basen-Engquist}. 2017.
\newblock \showarticletitle{Mobile health physical activity intervention
  preferences in cancer survivors: a qualitative study}.
\newblock {\em JMIR mHealth and uHealth\/} {5}, 1 (2017).
\newblock


\bibitem{warburton2006health}
{Darren~ER Warburton}, {Crystal~Whitney Nicol}, {and} {Shannon~SD Bredin}.
  2006.
\newblock \showarticletitle{Health benefits of physical activity: the
  evidence}.
\newblock {\em Canadian medical association journal\/} {174}, 6 (2006),
  801--809.
\newblock


\bibitem{xu2017analysis}
{Tongtong Xu}, {Ao Guo}, {and} {Jianhua Ma}. 2017.
\newblock \showarticletitle{Analysis of Temporal Features in Data Streams from
  Multiple Wearable Devices}. In {\em Cybernetics (CYBCONF), 2017 3rd IEEE
  International Conference on}. IEEE, 1--6.
\newblock


\end{thebibliography}

\end{document}